\documentclass[aps,prd,amsmath,amssymb,superscriptaddress,preprintnumbers,twocolumn,nofootinbib,10pt]{revtex4-1}
\pdfoutput=1
\usepackage{graphicx}
\usepackage{dcolumn}
\usepackage{bm}
\usepackage{amssymb}
\usepackage{latexsym}
\usepackage{booktabs}
\usepackage{amsmath}
\usepackage{multirow}
\usepackage{url}
\usepackage{footnote}
\usepackage{threeparttable}
\usepackage[bottom]{footmisc}

\usepackage[normalem]{ulem}
\usepackage{color}
\usepackage{array}
\usepackage{enumerate}
\usepackage{adjustbox}
\usepackage{makecell}
\usepackage{diagbox}
\usepackage{epstopdf}
\usepackage{epsfig}
\usepackage{longtable}
\usepackage{supertabular}
\usepackage{algorithm}
\usepackage{pifont}
\usepackage{algorithmic}
\usepackage{changepage}
\usepackage{setspace}
\usepackage{tabularx}
\newcolumntype{C}{>{\centering\arraybackslash}X}
\usepackage[colorlinks=true, linkcolor=blue, citecolor=blue]{hyperref}

\begin{document}

\title{Neutrino mass constraints in interacting dark energy models after DESI DR2}

\author{Hui Li}  
\affiliation{Liaoning Key Laboratory of Cosmology and Astrophysics, College of Sciences, Northeastern University, Shenyang 110819, China}

\author{Guo-Hong Du} 
\affiliation{Liaoning Key Laboratory of Cosmology and Astrophysics, College of Sciences, Northeastern University, Shenyang 110819, China}

\author{Tian-Nuo Li} 
\affiliation{Liaoning Key Laboratory of Cosmology and Astrophysics, College of Sciences, Northeastern University, Shenyang 110819, China}

\author{Hai-Li Li}  
\affiliation{College of Sciences, Shenyang Institute of Engineering, Shenyang 110136, China}

\author{Lu Feng} 
\affiliation{College of Physical Science and Technology, Shenyang Normal University, Shenyang 110034, China}

\author{Jing-Fei Zhang} 
\affiliation{Liaoning Key Laboratory of Cosmology and Astrophysics, College of Sciences, Northeastern University, Shenyang 110819, China}

\author{Xin Zhang}\thanks{Corresponding author}\email{zhangxin@mail.neu.edu.cn}   
\affiliation{Liaoning Key Laboratory of Cosmology and Astrophysics, College of Sciences, Northeastern University, Shenyang 110819, China}
\affiliation{MOE Key Laboratory of Data Analytics and Optimization for Smart Industry, Northeastern University, Shenyang 110819, China}
\affiliation{National Frontiers Science Center for Industrial Intelligence and Systems Optimization, Northeastern University, Shenyang 110819, China}

\begin{abstract}
Recent DESI observations indicate a deviation from the $\Lambda$CDM model, showing a preference for dynamical dark energy and thereby relaxing the upper limit on the neutrino mass within this framework. This deviation can also be explained by the presence of an interaction between dark energy and dark matter. In this work, we investigate the cosmological upper bounds on the total neutrino mass ($\sum m_{\nu}$) across four different interacting dark energy (IDE) models. The present analysis employs the latest DESI baryon acoustic oscillation, cosmic microwave background, and type Ia supernova datasets.  These results demonstrate that the upper bounds on $\sum m_{\nu}$ exhibit profound sensitivity to the specific phenomenological formulation of the interaction term. While the I$\Lambda$CDM2 model ($Q \propto H \rho_{\mathrm{c}}$) substantially relaxes the stringent upper limit ($\sum m_{\nu} < 0.129$ eV at 95\% confidence level), notably the I$\Lambda$CDM3 model ($Q \propto H_0 \rho_{\mathrm{de}}$), severely compresses the allowed parameter space, yielding a highly restrictive bound of $\sum m_{\nu} < 0.051$ eV. Furthermore, rigorous goodness-of-fit evaluations utilizing the Deviance Information Criterion and $\Delta\chi^2_{\mathrm{MAP}}$ indicate that the current observational data statistically favor these mass-suppressing IDE models. This establishes an exacerbated statistical tension between the observationally preferred IDE scenarios and the normal hierarchy lower bound ($\sim 0.06$ eV) determined by terrestrial neutrino oscillation experiments.
\end{abstract}

\maketitle
\section{Introduction}\label{sec1}

Recent observational data from the Dark Energy Spectroscopic Instrument (DESI) have brought new breakthroughs to cosmological research. In particular, the recently released DESI baryon acoustic oscillations (BAO) data, when analyzed in combination with cosmic microwave background (CMB) and type Ia supernova (SN) data, exhibit a significant deviation from the standard $\Lambda$CDM model~\cite{DESI:2025zgx,DESI:2024mwx}. This deviation achieves substantial statistical significance (approximately $2.8-4.2\sigma$) within the Chevallier-Polarski-Linder (CPL) parametrization, which implies that the equation of state (EoS) of dark energy may not be the cosmological constant ($w=-1$). Instead, it appears to exhibit a dynamical evolution, specifically evolving from $w < -1$ at early times to $w > -1$ at late times, which is referred to as quintom behavior~\cite{Feng:2004ad,Guo:2004fq,Zhang:2005kj,Zhang:2005yz,Zhang:2006qu}. This phenomenon, revealed by the DESI data, has rapidly sparked widespread interest and intense discussion within the cosmological community~\cite{Amendola:1999er,Wang:2016lxa,Cortes:2024lgw,Shlivko:2024llw,Luongo:2024fww,Wang:2024dka,Tada:2024znt,Bhattacharya:2024hep,Ramadan:2024kmn,Giare:2024gpk,Jiang:2024xnu,Yang:2024kdo,DESI:2025fii,DESI:2025wyn,deCruzPerez:2025dni,Gialamas:2024lyw,Ye:2024ywg,DESI:2024kob,Malekjani:2024bgi,Reboucas:2024smm,Park:2024pew,Li:2024qus,Wolf:2025jed,Shajib:2025tpd,Giare:2025pzu,Chaussidon:2025npr,Pang:2025lvh,RoyChoudhury:2025dhe,Paliathanasis:2025cuc,Scherer:2025esj,Giare:2024oil,Liu:2025mub,Teixeira:2025czm,Specogna:2025guo,Cheng:2025lod,Herold:2025hkb,Ozulker:2025ehg,Ormondroyd:2025iaf,Silva:2025twg,Ishak:2025cay,Zhang:2025lam,Cheng:2025hug,Cai:2025mas,Li:2025ops,Lee:2025pzo,Wu:2025vfs,Santos:2025wiv,Li:2025vuh,Li:2025dwz,Khoury:2025txd,Li:2025ula,Kessler:2025kju,Smith:2025icl,CosmoVerseNetwork:2025alb,Zhang:2025lam,Fazzari:2025lzd,Song:2025bio,Bella:2026zuk,Comini:2026nsj,Gonzalez-Fuentes:2026rgu,Pedrotti:2025ccw,Ren:2026jyw,Li:2026hwq,Yao:2025twv,Qiu:2025oop,Wang:2026kbg}. Although early perspectives questioned whether this signal might stem from systematic errors in the SN data~\cite{Huang:2025som,Efstathiou:2024xcq}, the rigorously calibrated DES-Dovekie SN sample subsequently released by the Dark Energy Survey collaboration retained $3.2\sigma$ evidence supporting this deviation, thus further consolidating the observational basis for the existence of dynamical dark energy~\cite{DES:2025sig}.

Additionally, neutrinos, as a unique relativistic component of the universe, play a non-negligible role in the cosmic expansion history and the formation of large-scale structures at late times~\cite{Zhang:2015rha,Allison:2015qca,Geng:2015haa,Chen:2015oga,Zhang:2015uhk,Huang:2015wrx,Chen:2016eyp,Moresco:2016nqq,Giusarma:2016phn,Lu:2016hsd,Wang:2016tsz,Kumar:2016zpg,Zhao:2016ecj,Vagnozzi:2017ovm,Guo:2017hea,Zhang:2017rbg,Yang:2017amu,Lorenz:2017fgo,Feng:2017nss,Xu:2016ddc,Li:2017iur,Wang:2017htc,Zhao:2017jma,Vagnozzi:2018jhn,Wang:2018lun,Giusarma:2018jei,Guo:2018gyo,Loureiro:2018pdz,Zhao:2018fjj,Feng:2019mym,Zhang:2019ipd,Feng:2019jqa,Liu:2020vgn,Yang:2020tax,Zhang:2020mox,Li:2020gtk,Yang:2020ope,Jin:2022tdf,Tanseri:2022zfe,Reeves:2022aoi,Pang:2023joc,Feng:2024lzh,RoyChoudhury:2024wri,Du:2024pai,Elbers:2025vlz,Zhou:2025nkb,Nair:2025uyn,Chudaykin:2025lww,DOnofrio:2025cuk,Sharma:2025iux,Du:2025xes,Giare:2025ath,DESI:2025ffm,RoyChoudhury:2025dhe,Jiang:2024viw,Ivanov:2026dvl,FrancoAbellan:2026ori,Ladeira:2026jne,Barua:2025adv,Yang:2026yaq}. Consequently, cosmological observations have become one of the most sensitive probes for limiting the absolute mass of neutrinos. While particle physics $\beta$-decay experiments provide direct laboratory measurements of $\sum m_{\nu}$ (with the latest KATRIN results yielding $\sum m_{\nu} < 1.35$~eV~\cite{KATRIN:2021uub,KATRIN:2024cdt}), the joint analysis of the CMB and DESI Data Release 2 (DR2) data within the $\Lambda$CDM model sets a cosmological upper bound on the total neutrino mass of $\sum m_{\nu} <$ 0.064 eV. This constraint not only far exceeds laboratory precision but also approaches the lower bound given by neutrino oscillation experiments and is in tension with the inverted hierarchy. This highlights that current cosmological observations impose significantly stricter limits on $\sum m_{\nu}$ than particle physics experiments~\cite{DESI:2025zgx}. 

Nevertheless, cosmological bounds on $\sum m_{\nu}$ exhibit strong model dependence, as different dark energy scenarios can substantially alter these constraints~\cite{Zhang:2014nta,Zhang:2015rha,Zhang:2017rbg,Feng:2017mfs,Guo:2018gyo,Feng:2017usu,Feng:2019jqa,Du:2024pai}. In a series of previous studies, we specifically focused on the cosmological degeneracy between $\sum m_{\nu}$ and the dark energy EoS parameter $w$~\cite{Zhang:2015uhk,Zhang:2017rbg,Zhao:2016ecj}. Our research demonstrated that the dynamical evolution of dark energy from $w<-1$ to $w>-1$, as hinted by DESI observations, can effectively relax the upper bound on the total neutrino mass \cite{Du:2024pai}. Building on this framework, \citet{Du:2025xes} demonstrated that incorporating weak gravitational lensing data to simultaneously constrain $\sum m_{\nu}$ and the effective number of relativistic species $N_{\rm eff}$ within the CPL model yields a $2.7\sigma$ measurement of a positive neutrino mass ($\sum m_\nu = 0.098^{+0.016}_{-0.037}\,{\rm eV}$).

However, purely dynamical dark energy models primarily alter the background expansion history of the universe, having a relatively limited direct impact on cosmic structure formation. Meanwhile, cosmological constraints on neutrino masses rely heavily on the CMB power spectrum, meaning that ncosmological constraints on neutrino mass are intrinsically much more sensitive to the details of structure formation than to background expansion alone. To further investigate this mechanism, we turn to explore the interacting dark energy (IDE) model~\cite{Zhang:2005rj,Zhang:2005rg,Fu:2011ab,Zhang:2017ize,DiValentino:2017iww,Yang:2018euj,DiValentino:2019ffd,Lucca:2021dxo,Gao:2021xnk,Wang:2021kxc,Giare:2024smz,Li:2024qso,Wang:2024vmw,Yang:2025uyv,Li:2026xaz,Pan:2025qwy}. Evaluated against the latest observational data, IDE models not only yield strong evidence for a non-zero dark sector interaction at a statistical confidence level of $3-5\sigma$, but their parameter constraints also significantly deviate from the standard cosmological limits~\cite{Li:2024qso,Giare:2024smz,Li:2025owk,Pan:2025qwy,Li:2026xaz,Wang:2025znm}. The most compelling feature of the IDE framework is that, while it can phenomenologically mimic the background expansion history of dynamical dark energy, its evolutionary dynamics at the perturbation level (which govern structure formation) are profoundly different. Therefore, reinvestigating neutrino mass constraints under the IDE framework is theoretically well-motivated and highly necessary.

In this work, we comprehensively investigate the impact of IDE models on neutrino mass measurements. We measure the total neutrino mass within the framework of IDE by examining four distinct forms of the interaction term $Q$. To constrain the models, we utilize a robust combination of the latest cosmological observations, incorporating BAO from DESI, CMB data from Planck and Atacama Cosmology Telescope (ACT), and SN compilations including PantheonPlus, DESY5, and DES-Dovekie.

This paper is organized as follows. In Sec.~\ref{sec2}, we briefly introduce the IDE models, massive neutrinos, and the observational datasets used in this work. In Sec.~\ref{sec3}, we report the joint constraint results and provide relevant discussions. Finally, we conclude this work in Sec.~\ref{sec4}.

\section{Methodology and Data}\label{sec2}
In the IDE scenario, we introduce a non-gravitational coupling between the dark energy and cold dark matter. The energy conservation equations for dark energy and cold dark matter are modified as
\begin{align}
\rho^{\prime}_{\mathrm{de}} + 3\mathcal{H}(1+w)\rho_{\mathrm{de}} &= a Q, \label{eq:rho_de} \\
\rho^{\prime}_{\mathrm{c}} + 3\mathcal{H}\rho_{\mathrm{c}} &= -a Q, \label{eq:rho_c}
\end{align}
where the prime denotes the derivative with respect to conformal time $\eta$, $\mathcal{H} = aH$ is the conformal Hubble parameter, and $w = p_{\mathrm{de}}/\rho_{\mathrm{de}}$ is the EoS parameter of dark energy. In this work, we focus on the I$\Lambda$CDM scenario where $w=-1$. $Q$ represents the energy transfer rate; a positive $Q$ indicates energy transfer from cold dark matter to dark energy.

The specific form of the interaction term $Q$ determines the dynamics of the model. We consider four typical phenomenological interaction models, which can be expressed mathematically as
\begin{align}
\text{I}\Lambda\text{CDM1}: \quad Q_1 &= \beta H \rho_{\mathrm{de}}, \label{eq:Q1} \\
\text{I}\Lambda\text{CDM2}: \quad Q_2 &= \beta H \rho_{\mathrm{c}}, \label{eq:Q2} \\
\text{I}\Lambda\text{CDM3}: \quad Q_3 &= \beta H_0 \rho_{\mathrm{de}}, \label{eq:Q3} \\
\text{I}\Lambda\text{CDM4}: \quad Q_4 &= \beta H_0 \rho_{\mathrm{c}}, \label{eq:Q4}
\end{align}
where $\beta$ is a dimensionless coupling constant, and $H_0$ is the Hubble constant. Given the lack of a fundamental microphysical theory describing the coupling between the dark sectors, a phenomenological approach is adopted, assuming the interaction term $Q$ is proportional to the energy densities~\cite{Amendola:1999qq,Billyard:2000bh}. To satisfy dimensional consistency, this term must be multiplied by a factor with units of inverse time. The dynamical Hubble parameter $H$ is a natural choice for this scale, as it facilitates obtaining analytical solutions to the background evolution equations. However, one might argue that local particle interactions should be independent of the global cosmic expansion. To address this concern, models scaled by the constant $H_0$ are considered, where $H_0$ is introduced solely to maintain dimensional validity without linking the interaction strength to the expansion history~\cite{Valiviita:2008iv,Boehmer:2008av,Caldera-Cabral:2008yyo,He:2008si,Clemson:2011an}.

To verify the viability of these models against observational data, we must consider the evolution of cosmological perturbations. It is well known that the interaction terms in the perturbation equations often involve factors inversely proportional to $1+w$. Consequently, when $w \approx -1$, the dark energy perturbations can suffer from large-scale instabilities in IDE models~\cite{Majerotto:2009zz,Clemson:2011an}. To mitigate these nonphysical divergences, the extended Parameterized Post-Friedmann approach is implemented, which provides a stable calculation of dark energy perturbations across the entire parameter space~\cite{Li:2014cee,Li:2015vla,Zhang:2017ize,Li:2023fdk}.

We consider the contribution of massive neutrinos to the cosmic energy budget. The total energy density of massive neutrinos, $\rho_{\nu}$, is determined by the phase space distribution of neutrino species~\cite{WMAP:2010qai}. Using the comoving momentum $q$, this can be expressed as
\begin{equation}
\rho_{\nu}(a) = \frac{a^{-4}}{\pi^{2}} \int \frac{q^{2}dq}{e^{q/T_{\nu 0}} + 1} \sum_{i} \sqrt{q^{2} + m_{{i}}^{2}a^{2}},
\label{eq:rho_nu_integral}
\end{equation}
where $T_{\nu 0} = (4/11)^{1/3}T_{\mathrm{cmb}} \approx 1.945$ K is the present-day neutrino temperature, and $m_{i}$ represents the mass of the $i$-th neutrino eigenstate~\cite{WMAP:2010qai}. In the universe, the total relativistic energy density of radiation is given by
\begin{equation}
\rho_{\mathrm{r}} = \left[1 + N_{\mathrm{eff}} \frac{7}{8} \left(\frac{4}{11}\right)^{\frac{4}{3}}\right]\rho_{{\gamma}},
\label{eq:rho_r}
\end{equation}
where $\rho_{{\gamma}}$ is the photon energy density~\cite{WMAP:2010qai}. The standard cosmological model has $N_{\mathrm{eff}} = 3.044$~\cite{Akita:2020szl,Froustey:2020mcq,Bennett:2020zkv}. At late times ($a \to \infty$), neutrinos become non-relativistic and contribute to the matter component of the universe. In this limit, Eq. \eqref{eq:rho_nu_integral} asymptotically approaches~\cite{WMAP:2010qai}
\begin{equation}
\rho_{\nu}(a) \to \frac{\sum m_{\nu}}{93.14 h^{2} \text{ eV}} \rho_{\mathrm{crit},0} a^{-3},
\label{eq:rho_nu_late}
\end{equation}
where $\rho_{\mathrm{crit},0} = 3H_{0}^2 / (8\pi G)$ is the critical density of the universe at the present time, $h\equiv H_0/(100\,\rm km\,s^{-1}\, Mpc^{-1})$ is the dimensionless Hubble constant.

By incorporating $\sum m_{\nu}$ into the baseline interactive framework, the extended models are hereafter denoted as
the I$\Lambda$CDM+$\sum m_{\nu}$ models. In this analysis, we adopt a hybrid scheme for the neutrino mass hierarchy based on $\sum m_\nu$ calculation. Specifically, when $\sum m_{\nu} < 0.06$ eV, we perform the calculation assuming a degenerate hierarchy (DH), where the three active neutrino eigenstates are treated as equal ($m_{1} = m_{2} = m_{3}$). Conversely, when $\sum m_{\nu} > 0.06$ eV, we calculate the physical quantities assuming a normal hierarchy (NH). We treat $\sum m_{\nu}$ as a free parameter with a flat prior $\sum m_{\nu} \in [0, 5]$ eV.

To constrain the free parameters of the IDE model and $\sum m_{\nu}$, we employ the public package \texttt{Cobaya}. The theoretical predictions are computed using \texttt{IDECAMB}~\cite{Li:2023fdk}. The MCMC chains are analyzed using the public package \texttt{GetDist}~\cite{Lewis:2019xzd}. The specific datasets utilized in this work are as follows:

\begin{itemize}
    \item \textbf{CMB}: We utilize the Planck CMB likelihoods for TT and EE spectra in the multipole range $2 < \ell < 30$~\cite{Planck:2018vyg,Planck:2019nip}. For high multipoles, we employ the NPIPE \texttt{CamSpec} likelihood for the TT spectrum in the range $30 \le \ell \le 2500$, and for the TE and EE spectra in the range $30 < \ell < 2000$~\cite{Efstathiou:2019mdh,Rosenberg:2022sdy}. Additionally, we include the CMB lensing likelihood utilizing the latest high-precision reconstruction from NPIPE PR4 Planck data~\cite{Carron:2022eyg} and ACT Data Release 6~\cite{ACT:2023dou,ACT:2023kun,AtacamaCosmologyTelescope:2025blo}.

    \item \textbf{DESI}: We employ the BAO measurements from the DESI DR2~\cite{DESI:2025zgx}. The data vector includes the transverse comoving distance $D_{\mathrm{M}}/r_{\mathrm{d}}$, the angle-averaged distance $D_{\mathrm{V}}/r_{\mathrm{d}}$, and the Hubble horizon $D_{\mathrm{H}}/r_{\mathrm{d}}$. These measurements are derived from multiple tracers, including the bright galaxy sample, luminous red galaxies, emission line galaxies, quasars, and the Lyman-$\alpha$ forest.

    \item \textbf{DESY5}: The DESY5 sample comprises 1829 supernovae. It consists of 1635 photometrically classified supernovae from the Dark Energy Survey 5-year data in the redshift range $0.1 < z < 1.3$, and 194 low-redshift supernovae from external samples (CfA3, CfA4, CSP, and Foundation) in the range $0.025 < z < 0.1$~\cite{Hicken:2009df,Hicken:2012zr,Krisciunas:2017yoe,Foley:2017zdq,DES:2024jxu}.

    \item \textbf{DES-Dovekie}: We utilize the fully recalibrated DESY5 sample, referred to as DES-Dovekie. This dataset incorporates improved photometric cross-calibration using Gaia and HST white dwarf standards to reduce systematic uncertainties in the color-luminosity relation. The sample comprises $1623$ DES supernovae and $197$ low-z supernovae, providing updated constraints on the distance modulus~\cite{DES:2025sig}.
    
    \item \textbf{PantheonPlus}: The PantheonPlus sample comprises 1550 spectroscopically confirmed supernovae collected from 18 different surveys, spanning the redshift range $0.01 < z < 2.26$~\cite{Brout:2022vxf}.

\end{itemize}

\begin{table*}[tbp]
\centering
\caption{The $1\sigma$ confidence regions (or $2\sigma$ upper limits) of cosmological parameters obtained by the DESI, CMB, DESY5, PantheonPlus, and DES-Dovekie data for the $\Lambda$CDM+$\sum m_{\nu}$, I$\Lambda$CDM2+$\sum m_{\nu}$, I$\Lambda$CDM3+$\sum m_{\nu}$, and I$\Lambda$CDM4+$\sum m_{\nu}$ models. For the parameter $\sum m_{\nu}$, central values cannot be determined in most cases, and we provide the $2\sigma$ upper limits. Here, $H_0$ is in units of $\mathrm{km\,s^{-1}\,Mpc^{-1}}$ and $\sum m_\nu$ is in units of $\mathrm{eV}$.}
\label{table1}
\renewcommand{\arraystretch}{1.5}
\setlength{\tabcolsep}{5pt}
\begin{tabular*}{\textwidth}{@{\extracolsep{\fill}}llcccc}
\hline\hline 
Model & Parameter & CMB+DESI & +DES-Dovekie & +PantheonPlus & +DESY5 \\
\hline
\multirow{5}{*}{$\Lambda$CDM+$\sum m_{\nu}$} 
  & $H_0$ & $68.35 \pm 0.29$ & $68.22 \pm 0.30$ & $68.28^{+0.31}_{-0.28}$ & $68.15 \pm 0.29$ \\
  & $\Omega_{\mathrm{m}}$ & $0.3011 \pm 0.0037$ & $0.3027 \pm 0.0037$ & $0.3019^{+0.0034}_{-0.0040}$ & $0.3036 \pm 0.0037$ \\
  & $\sigma_8$ & $0.8167^{+0.0069}_{-0.0055}$ & $0.8162^{+0.0072}_{-0.0058}$ & $0.8169^{+0.0068}_{-0.0057}$ & $0.8161^{+0.0070}_{-0.0061}$ \\
  & $\beta$ & --- & --- & --- & --- \\
  & $\sum m_{\nu}$ & $< 0.065$ & $< 0.073$ & $< 0.066$ & $< 0.079$ \\
\hline
\multirow{5}{*}{I$\Lambda$CDM1+$\sum m_{\nu}$} 
  & $H_0$ & $69.11^{+0.91}_{-1.00}$ & $67.76^{+0.49}_{-0.56}$ & $67.84 \pm 0.54$ & $67.27 \pm 0.53$ \\
  & $\Omega_{\mathrm{m}}$ & $0.2730^{+0.0370}_{-0.0320}$ & $0.3220^{+0.0190}_{-0.0160}$ & $0.3190 \pm 0.0190$ & $0.3390^{+0.0190}_{-0.0170}$ \\
  & $\sigma_8$ & $0.8940^{+0.0580}_{-0.1100}$ & $0.7800^{+0.0280}_{-0.0380}$ & $0.7850^{+0.0310}_{-0.0400}$ & $0.7510^{+0.0260}_{-0.0350}$ \\
  & $\beta$ & $0.0970^{+0.1100}_{-0.1200}$ & $-0.0680^{+0.0580}_{-0.0680}$ & $-0.0590 \pm 0.0660$ & $-0.1270 \pm 0.0630$ \\
  & $\sum m_{\nu}$ & $< 0.083$ & $< 0.064$ & $< 0.063$ & $< 0.054$ \\
\hline
\multirow{5}{*}{I$\Lambda$CDM2+$\sum m_{\nu}$} 
  & $H_0$ & $68.78 \pm 0.39$ & $68.54 \pm 0.38$ & $68.59 \pm 0.37$ & $68.45 \pm 0.38$ \\
  & $\Omega_{\mathrm{m}}$ & $0.2963 \pm 0.0045$ & $0.2991 \pm 0.0044$ & $0.2986 \pm 0.0043$ & $0.3002 \pm 0.0044$ \\
  & $\sigma_8$ & $0.8250 \pm 0.0091$ & $0.8225 \pm 0.0089$ & $0.8220 \pm 0.0088$ & $0.8212 \pm 0.0089$ \\
  & $\beta$ & $0.0018 \pm 0.0010$ & $0.0015 \pm 0.0010$ & $0.0016^{+0.0009}_{-0.0011}$ & $0.0014 \pm 0.0010$ \\
  & $\sum m_{\nu}$ & $< 0.122$ & $< 0.129$ & $< 0.131$ & $< 0.130$ \\
\hline
\multirow{5}{*}{I$\Lambda$CDM3+$\sum m_{\nu}$} 
  & $H_0$ & $68.80 \pm 1.20$ & $67.46 \pm 0.56$ & $67.65^{+0.56}_{-0.63}$ & $66.97 \pm 0.54$ \\
  & $\Omega_{\mathrm{m}}$ & $0.2810 \pm 0.0510$ & $0.3400 \pm 0.0230$ & $0.3320^{+0.0270}_{-0.0240}$ & $0.3620^{+0.0240}_{-0.0220}$ \\
  & $\sigma_8$ & $0.8860^{+0.0900}_{-0.1600}$ & $0.7460^{+0.0340}_{-0.0440}$ & $0.7610^{+0.0350}_{-0.0540}$ & $0.7120^{+0.0280}_{-0.0410}$ \\
  & $\beta$ & $0.080^{+0.240}_{-0.200}$ & $-0.180 \pm 0.110$ & $-0.140 \pm 0.120$ & $-0.270 \pm 0.110$ \\
  & $\sum m_{\nu}$ & $< 0.071$ & $< 0.059$ & $< 0.062$ & $< 0.051$ \\
\hline
\multirow{5}{*}{I$\Lambda$CDM4+$\sum m_{\nu}$} 
  & $H_0$ & $68.95 \pm 0.63$ & $68.11 \pm 0.52$ & $68.30 \pm 0.54$ & $67.80 \pm 0.52$ \\
  & $\Omega_{\mathrm{m}}$ & $0.2880^{+0.0120}_{-0.0130}$ & $0.3052^{+0.0098}_{-0.0110}$ & $0.3010 \pm 0.0110$ & $0.3120 \pm 0.0110$ \\
  & $\sigma_8$ & $0.8310 \pm 0.0150$ & $0.8130 \pm 0.0130$ & $0.8180 \pm 0.0130$ & $0.8070 \pm 0.0130$ \\
  & $\beta$ & $0.0420 \pm 0.0390$ & $-0.0080 \pm 0.0320$ & $0.0040 \pm 0.0340$ & $-0.0260 \pm 0.0330$ \\
  & $\sum m_{\nu}$ & $< 0.090$ & $< 0.073$ & $< 0.079$ & $< 0.071$ \\
\hline 
\hline
\end{tabular*}
\end{table*}

\section{Results and Discussions}\label{sec3}

\begin{figure*}[tbp]
\centering
\includegraphics[width=1\textwidth]{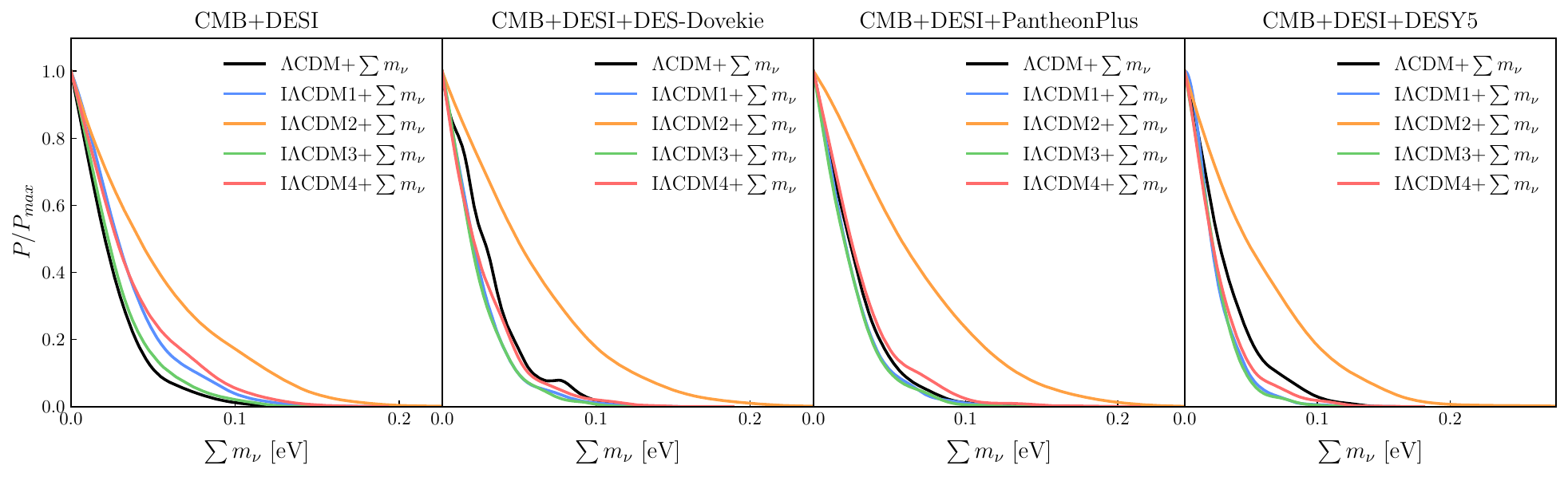} 
\caption{The 1D marginalized posterior constraints on $\sum m_{\nu}$ using the CMB+DESI, CMB+DESI+DES-Dovekie, CMB+DESI+PantheonPlus, and CMB+DESI+DESY5 datasets. The black curve represents the standard $\Lambda$CDM+$\sum m_{\nu}$ model, while the colored curves correspond to different IDE models. }
\label{fig1}
\end{figure*}

\begin{figure*}[tbp]
\centering
\includegraphics[width=1\textwidth]{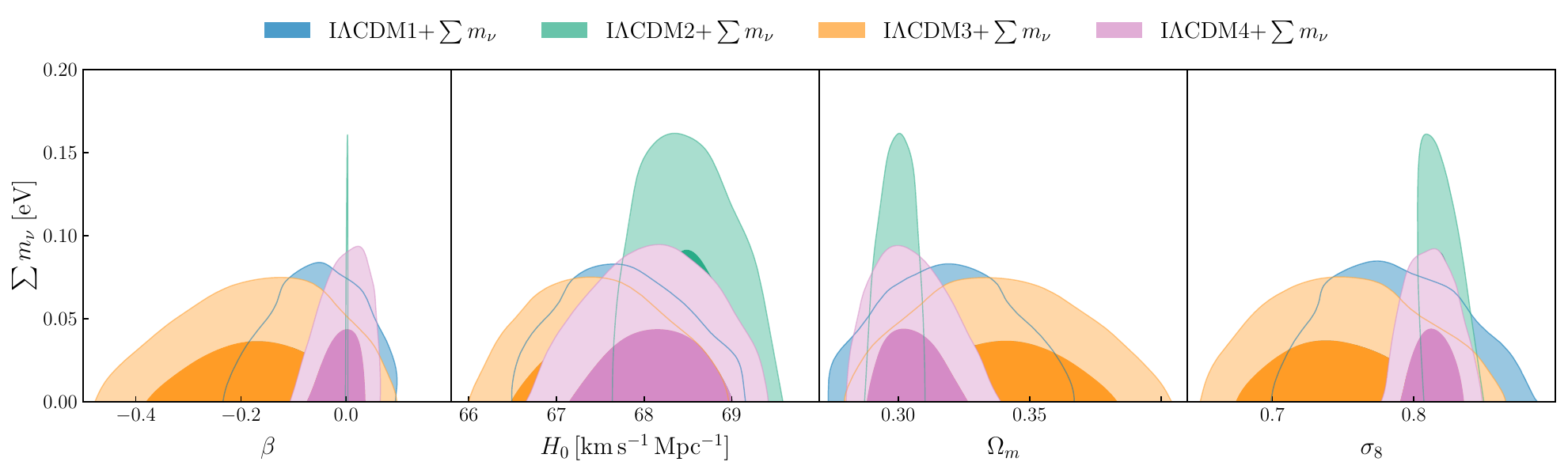} 
\caption{A comparison of the 2D marginalized contours of $\sum m_{\nu}$ with $\beta$, $H_0$, $\Omega_\mathrm{m}$, and $\sigma_8$ using CMB+DESI+DES-Dovekie data.}
\label{fig2}
\end{figure*}

\begin{figure*}[tbp]
\centering
\includegraphics[width=1\textwidth]{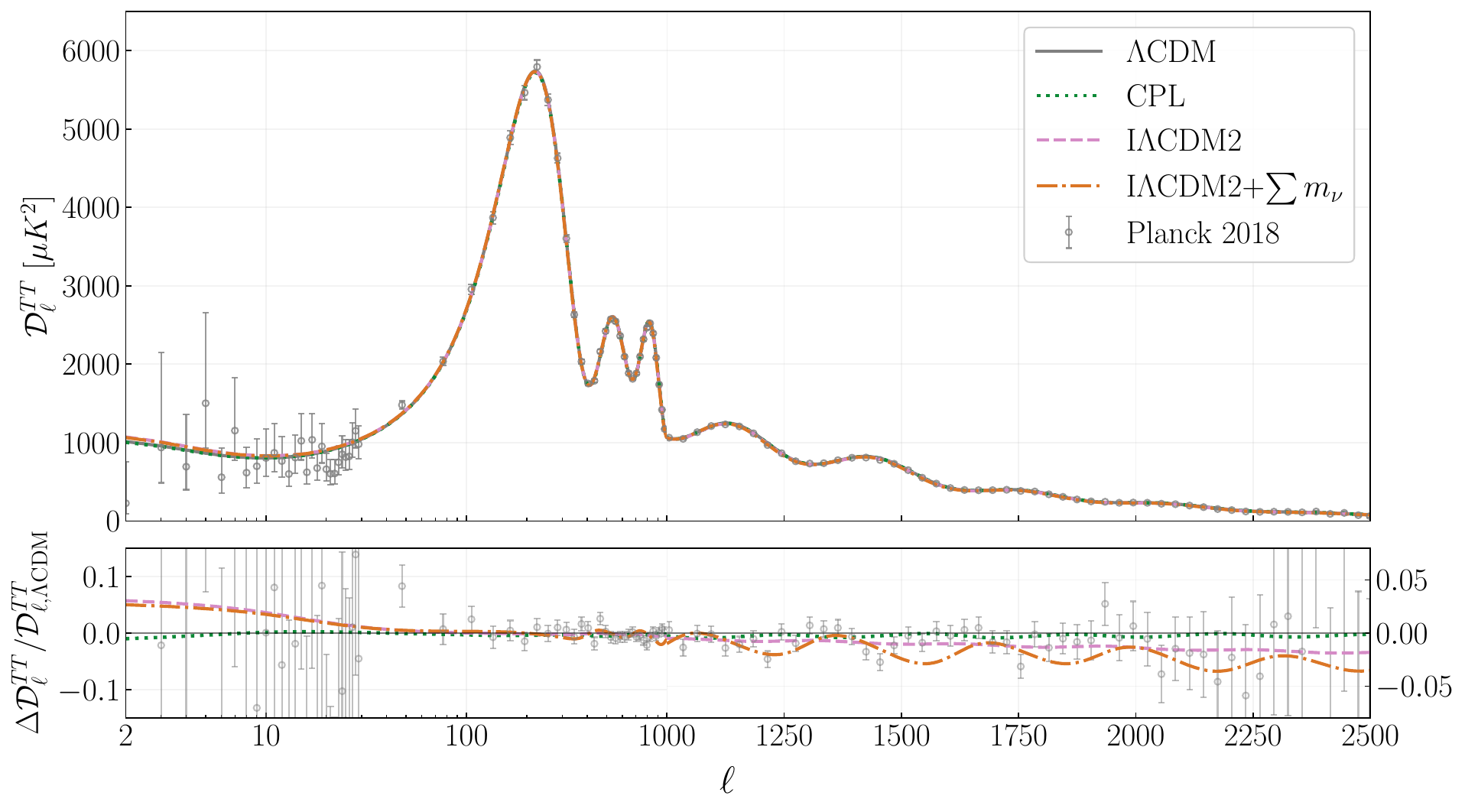} 
\caption{The best-fit CMB TT power spectra obtained from the CMB+DESI+DES-Dovekie data. We compare the theoretical prediction curves of the $\Lambda$CDM model with those of the CPL, I$\Lambda$CDM2, and I$\Lambda$CDM2+$\sum m_{\nu}$ models, superimposed with the Planck 2018 observational data points and their error bars. We also display the relative fractional differences of the extended models with respect to the $\Lambda$CDM model.}
\label{fig3}
\end{figure*}

In this section, we present the parameter constraints derived from various data combinations; these results are summarized in Table~\ref{table1}. To visualize the statistical distribution properties of key parameters, the one-dimensional (1D) marginalized posterior distributions and two-dimensional (2D) contour plots are displayed in Figs.~\ref{fig1} and \ref{fig2}, respectively. Additionally, Figure~\ref{fig3} illustrates the fitting results for the Planck 2018 CMB TT power spectrum. Furthermore, to provide a more quantitative comparison of the goodness-of-fit across models, we calculate the total $\Delta\chi^2_{\mathrm{MAP}}$ for each model relative to the $\Lambda\mathrm{CDM}+\sum m_{\nu}$ baseline, as well as the Deviance Information Criterion (DIC), the details of which are provided in Table~\ref{table2}.

Figure~\ref{fig1} illustrates the constraints on $\sum m_{\nu}$ under different models and dataset combinations. First, when using only CMB+DESI data, the $\Lambda$CDM+$\sum m_{\nu}$ model yields a constraint of $\sum m_{\nu} < 0.065$ eV. In comparison, the four IDE models relax the constraints on $\sum m_{\nu}$ to varying degrees: $\sum m_{\nu} < 0.083$ eV for I$\Lambda$CDM1+$\sum m_{\nu}$, $\sum m_{\nu} < 0.071$ eV for I$\Lambda$CDM3+$\sum m_{\nu}$, and $\sum m_{\nu} < 0.090$ eV for I$\Lambda$CDM4+$\sum m_{\nu}$ show smaller relaxations, while I$\Lambda$CDM2+$\sum m_{\nu}$ exhibits the most significant relaxation, with the upper limit extending to $\sum m_{\nu} < 0.122$ eV, an increase of 88\%. However, using the CMB+DESI+PantheonPlus data, the upper limits for I$\Lambda$CDM2+$\sum m_{\nu}$ and I$\Lambda$CDM4+$\sum m_{\nu}$ are relaxed, whereas those for I$\Lambda$CDM1+$\sum m_{\nu}$ and I$\Lambda$CDM3+$\sum m_{\nu}$ are suppressed. With the CMB+DESI+DES-Dovekie and CMB+DESI+DESY5 datasets, only I$\Lambda$CDM2+$\sum m_{\nu}$ relaxes the upper limit of $\sum m_{\nu}$, while the limits for the remaining models are suppressed to varying degrees. Taking CMB+DESI+DES-Dovekie as an example, I$\Lambda$CDM2+$\sum m_{\nu}$ yields an upper limit of $\sum m_{\nu} < 0.129$ eV, a relaxation of approximately 76\% compared to the $\Lambda$CDM+$\sum m_{\nu}$ model, while the other models provide stricter constraints.

The physical origin of this difference lies in the sign of the coupling parameter $\beta$ and its degeneracy with $\sum m_{\nu}$. The CMB+DESI data tend to favor a positive $\beta$ value; however, after including DES-Dovekie, the $\beta$ values for all models except I$\Lambda$CDM2+$\sum m_{\nu}$ become negative. Due to the parameter degeneracy between $\beta$ and $\sum m_{\nu}$, when $\beta < 0$ (dark energy decays into dark matter), $\sum m_{\nu}$ upper limit is suppressed; conversely, when $\beta > 0$ (dark matter decays into dark energy), the limit is relaxed. 

It is worth noting that under the combined dataset of CMB+DESI+DESY5, the upper limit given by the I$\Lambda$CDM1+$\sum m_{\nu}$ model is $\sum m_{\nu} < 0.054$~eV, while the I$\Lambda$CDM3+$\sum m_{\nu}$ model yields $\sum m_{\nu} < 0.051$~eV. In models where the interaction term $Q$ is proportional to $\rho_{\mathrm{de}}$, $\sum m_{\nu}$ remains less than $0.06$~eV at the $95\%$ confidence level. This result is in conflict with the minimum mass lower bound under the NH provided by particle physics neutrino oscillation experiments. Recent DESI studies have also found that if the effective $\sum m_{\nu}$ is allowed to take negative values in the fit, the peak of the likelihood function for the DESI+CMB data falls in the region of $\sum m_{\nu,\mathrm{eff}} < 0$. In this case, the conflict between the cosmological results and the minimum mass lower bound given by terrestrial experiments increases to $3.0\sigma$. These results collectively imply that in current cosmological measurements, even when adopting models preferred by current data, there is still a tension between $\sum m_{\nu}$ and particle physics measurement results.

Recent studies have discussed that this tension might originate from the clustering behavior of cosmic matter. Since both $\Omega_{\rm{m}}$ and massive neutrinos affect the evolution of large-scale structures, the growth index ($\gamma$) is introduced to characterize the growth rate of matter perturbations, with the standard $\Lambda$CDM model predicting $\gamma \simeq 0.55$. If $\gamma$ is treated as a free parameter, the upper bound on $\sum m_{\nu}$ can be relaxed to $0.13 - 0.2$~eV. This mechanism can eliminate the conflict without changing the fitted value of $\Omega_{\rm{m}}$, but it requires the growth rate of cosmic matter structures to be suppressed: to satisfy the lower bound of terrestrial experiments ($\sum m_\nu \ge 0.06$~eV), it is required that $\gamma > 0.55$. After introducing this physical lower bound prior, current data exhibits a statistical preference of $2.5\sigma - 3.0\sigma$ for $\gamma > 0.55$.

To more clearly illustrate the correlations between the neutrino mass and other parameters, we present the two-dimensional contour plots of $\sum m_{\nu}$ versus key parameters ($\beta$, $H_0$, $\Omega_\mathrm{m}$, and $\sigma_8$) for each model using the CMB+DESI+DES-Dovekie data in Fig.~\ref{fig2}. We can clearly observe that the different forms of interaction significantly alter the correlations between parameters. Among the four models, I$\Lambda$CDM2+$\sum m_{\nu}$ is subject to the strictest constraints. In the matter-dominated epoch of the early universe, both the Hubble parameter $H$ and the critical density $\rho_{\mathrm{c}}$ are at extremely high values, implying that the interaction source term $H \rho_{\mathrm{c}}$ itself is large. However, the observational precision of the CMB is extremely high, and any substantial unsuppressed energy exchange would induce severe deviations, disrupting the theoretical concordance with observational data. Consequently, to maintain the calculated interaction intensity within CMB constraints despite the extremely large source term $H \rho_{\mathrm{c}}$, $\beta$ is forced to be suppressed to a very small numerical range. In contrast, the other three models are less constrained. Specifically, the interaction terms in I$\Lambda$CDM1+$\sum m_{\nu}$ and I$\Lambda$CDM3+$\sum m_{\nu}$ depend on the dark energy density $\rho_{\mathrm{de}}$. Since dark energy only becomes dominant in the late universe ($z < 1$), the interaction term is minimal in the early universe, thereby preserving a larger allowable parameter space. On the other hand, for I$\Lambda$CDM4+$\sum m_{\nu}$ and I$\Lambda$CDM2+$\sum m_{\nu}$, although both involve the critical density $\rho_{\mathrm{c}}$, the I$\Lambda$CDM4+$\sum m_{\nu}$ formula uses the constant $H_0$ instead of the dynamically evolving $H(z)$. Since $H(z)$ was far greater than $H_0$ in the past, the interaction generated by the I$\Lambda$CDM4+$\sum m_{\nu}$ model in the early universe (at high redshift) is far smaller than that of I$\Lambda$CDM2+$\sum m_{\nu}$. Therefore, the influence of this model on early physical processes is relatively weak, ultimately resulting in relatively loose parameter constraints.

We also observe a negative correlation between $\sigma_8$ and $\sum m_{\nu}$. $\sigma_8$ directly reflects the degree of matter clustering. However, as a high-velocity matter component, neutrinos induce a free-streaming effect that smooths out small-scale structures in the cosmic matter distribution. The larger $\sum m_{\nu}$, the more significant its suppression of structural growth, manifesting as a suppression of the matter power spectrum at high wavenumbers (small scales). Since the scale corresponding to $\sigma_8$ falls precisely within the sensitive range of this effect, an increase in $\sum m_{\nu}$ inevitably leads to a decrease in the value of $\sigma_8$. It is worth noting that for models where $Q \propto \rho_{\mathrm{de}}$ (I$\Lambda$CDM1+$\sum m_{\nu}$ and I$\Lambda$CDM3+$\sum m_{\nu}$), even though the yielded $\sum m_{\nu}$ values are small, the $\sigma_8$ values remain low, a phenomenon consistent with results from other literature~\cite{Giare:2024gpk,Giare:2025ath,Li:2025muv,Li:2026xaz}.

To visually assess the goodness of fit of each model to the observational data, we present the best-fit curves for the Planck 2018 CMB TT power spectrum in Fig.~\ref{fig3}. We plot the best-fit theoretical curves derived from the combined CMB+DESI+DES-Dovekie dataset, along with the relative fractional differences of the extended models compared to the $\Lambda$CDM model. The models considered include CPL, I$\Lambda$CDM2, and I$\Lambda$CDM2+$\sum m_{\nu}$. As illustrated in the figure, the power spectrum of the CPL model is nearly indistinguishable from that of the $\Lambda$CDM model. This suggests that, under current constraints, the dynamical nature of dark energy has a negligible impact on the shape of the CMB power spectrum. This is attributed to the fact that the CPL model primarily modifies the expansion history at the background level.

In contrast, the interacting dark energy models exhibit distinct deviations. The I$\Lambda$CDM2 model shows a maximum enhancement of approximately 5\% at large scales ($\ell \lesssim 30$) and a visible suppression at small scales ($\ell \gtrsim 1000$) relative to $\Lambda$CDM. Notably, near the first acoustic peak, the observational data points fit the $\Lambda$CDM model extremely well with small relative errors, imposing stringent constraints on these interacting models. As previously mentioned, the interaction strength within these coupled models is tightly constrained in this region. Further examining the models incorporating massive neutrinos, where $\sum m_{\nu}$ is treated as a free parameter, we find that the neutrino mass primarily affects the small-scale power spectrum. The I$\Lambda$CDM2+$\sum m_{\nu}$ model exhibits a pronounced suppression effect at small scales; particularly in the high-$\ell$ region ($\ell > 1000$), the free-streaming effect of neutrinos leads to an oscillatory damping with a maximum suppression amplitude reaching nearly 6\%. In the higher $\ell$ region ($\ell > 2000$), although the neutrino extensions predict varying degrees of power spectrum suppression, the substantial increase in the error bars of the observational data points renders all considered models broadly compatible with current observations.

\begin{table}[htbp]
\centering
\caption{Summary of goodness-of-fit and model comparison metrics for the $\mathrm{I}\Lambda\mathrm{CDM}+\sum m_{\nu}$ models relative to $\Lambda\mathrm{CDM}+\sum m_{\nu}$. For each data combination we report the Total $\Delta\chi^2_{\text{MAP}}$, and the $\Delta$DIC. Negative values indicate an improvement with respect to $\Lambda\mathrm{CDM}+\sum m_{\nu}$.}
\small
\renewcommand{\arraystretch}{1.3}
\begin{tabular*}{\columnwidth}{@{\extracolsep{\fill}}l c c}
\hline
\hline
\multirow{1}{*}{Model/Data} & \multicolumn{1}{c}{$\Delta\chi^2_{\text{MAP}}$} & \multirow{1}{*}{$\Delta\mathrm{DIC}$} \\
\hline
\textbf{$\mathrm{I}\Lambda\mathrm{CDM}1+\sum m_{\nu}$} & & \\
CMB+DESI & $-0.27$ & $-6.59$ \\
CMB+DESI+PantheonPlus & $-1.45$ & $-0.07$ \\
CMB+DESI+DESY5 & $-4.12$ & $-3.29$ \\
CMB+DESI+DES-Dovekie & $-3.44$ & $-0.50$ \\
\hline
\textbf{$\mathrm{I}\Lambda\mathrm{CDM}2+\sum m_{\nu}$} & & \\
CMB+DESI & $-0.97$ & $-0.49$ \\
CMB+DESI+PantheonPlus & $0.73$ & $0.18$ \\
CMB+DESI+DESY5 & $0.17$ & $1.39$ \\
CMB+DESI+DES-Dovekie & $-0.20$ & $0.72$ \\
\hline
\textbf{$\mathrm{I}\Lambda\mathrm{CDM}3+\sum m_{\nu}$} & & \\
CMB+DESI & $0.57$ & $2.12$ \\
CMB+DESI+PantheonPlus & $-2.04$ & $-3.88$ \\
CMB+DESI+DESY5 & $-6.95$ & $-8.03$ \\
CMB+DESI+DES-Dovekie & $-2.92$ & $-3.58$ \\
\hline
\textbf{$\mathrm{I}\Lambda\mathrm{CDM}4+\sum m_{\nu}$} & & \\
CMB+DESI & $0.64$ & $1.05$ \\
CMB+DESI+PantheonPlus & $0.72$ & $1.83$ \\
CMB+DESI+DESY5 & $0.97$ & $1.06$ \\
CMB+DESI+DES-Dovekie & $0.55$ & $1.50$ \\
\hline
\hline
\end{tabular*}
\label{table2}
\end{table}

To provide a more quantitative comparison of the goodness-of-fit among different models, we calculated the total $\Delta\chi^2_{\mathrm{MAP}}$ for each model relative to the $\Lambda\mathrm{CDM}+\sum m_{\nu}$ baseline. This metric is defined as the difference in $\chi^2$ between the given model and the standard model, evaluated at the maximum a posteriori (MAP) estimate. Furthermore, we computed the DIC to provide a complementary assessment of model performance. 

The results indicate significant variations in how different models adapt to the data: under the $\mathrm{I}\Lambda\mathrm{CDM}3+\sum m_{\nu}$ model, the CMB+DESI dataset slightly favors the standard model; however, the inclusion of SN data shifts the preference towards the interacting model, notably yielding $\Delta\chi^2_{\mathrm{MAP}} = -6.95$ and $\Delta\mathrm{DIC} = -8.03$ for the CMB+DESI+DESY5 combination. In contrast, the $\mathrm{I}\Lambda\mathrm{CDM}2+\sum m_{\nu}$ model demonstrates an advantage only with the CMB+DESI data, which vanishes upon the inclusion of additional datasets. 

Notably, the $\mathrm{I}\Lambda\mathrm{CDM}1+\sum m_{\nu}$ model exhibits the most robust fitting performance, yielding negative values for both $\Delta\chi^2_{\mathrm{MAP}}$ and $\Delta\mathrm{DIC}$ across all data combinations. Conversely, the $\mathrm{I}\Lambda\mathrm{CDM}4+\sum m_{\nu}$ model presents positive metric values across all combinations, implying that the introduction of this model fails to improve the goodness-of-fit, with the data consistently favoring the standard $\Lambda\mathrm{CDM}+\sum m_{\nu}$ model.

\section{Conclusion}\label{sec4}

In this work, we investigate the cosmological constraints on $\sum m_{\nu}$ within the IDE framework, utilizing the latest observational data from CMB, DESI DR2, and SN (DESY5, PantheonPlus, and DES-Dovekie). We focus on exploring four distinct forms of interaction models. Furthermore, we perform a quantitative comparison of the goodness-of-fit across different models by calculating $\Delta\chi^2_{\mathrm{MAP}}$ and the DIC.

Our analysis indicates that introducing an interaction significantly alters the constraints on $\sum m_{\nu}$, with the specific effects depending strongly on the form of the interaction term $Q$. We find that the $\mathrm{I}\Lambda\mathrm{CDM}2+\sum m_{\nu}$ model ($Q \propto H \rho_{\mathrm{c}}$) yields the most relaxed constraints on $\sum m_{\nu}$. Compared to the standard $\Lambda$CDM model, this model relaxes the upper limit of $\sum m_{\nu}$ by approximately $64\%$--$98\%$ across different data combinations, relaxing the constraint to $\sum m_{\nu} < 0.129$ eV. This is primarily attributed to the positive values of the coupling constant $\beta$ in this model, which accommodates a larger $\sum m_{\nu}$. Conversely, for models where the interaction term is proportional to the dark energy density, specifically $\mathrm{I}\Lambda\mathrm{CDM}1+\sum m_{\nu}$ and $\mathrm{I}\Lambda\mathrm{CDM}3+\sum m_{\nu}$, dark energy dominates only at late times. Because these models involve the decay of dark energy into dark matter, they are less constrained in the early universe, leading to a further suppression of $\sum m_{\nu}$ upper limit. Notably, in the $\mathrm{I}\Lambda\mathrm{CDM}3+\sum m_{\nu}$ model ($Q \propto H_0 \rho_{\mathrm{de}}$), the constraint obtained by combining with DESY5 data is $\sum m_{\nu} < 0.051$ eV. This result suppresses $\sum m_{\nu}$ below the minimum mass bound of the NH. In particular, the quantitative comparison of model goodness-of-fit reveals that when DESY5 SN data are included, the $\mathrm{I}\Lambda\mathrm{CDM}3+\sum m_{\nu}$ model exhibits a relatively strong preference over the standard $\Lambda$CDM model ($\Delta\mathrm{DIC} = -8.03$). This result further underscores the possible tension between current cosmological observations and the NH lower bound from particle physics neutrino oscillation experiments.

In summary, our study highlights the pivotal impact of IDE models on $\sum m_{\nu}$ measurements and further emphasizes the strong dependence of cosmological $\sum m_{\nu}$ constraints on the dark energy model. It is worth noting that within the framework of the IDE models favored by current observational data, the derived upper limit on $\sum m_{\nu}$ is lower than the lower bound required by particle physics experiments, revealing a tension between cosmological observations and particle physics measurements. In the future, with the release of full-shape power spectrum data from the complete 5-year DESI observations, as well as the arrival of more precise observational data from Euclid~\cite{Euclid:2024yrr} and the Large Synoptic Survey Telescope~\cite{LSST:2008ijt}, we will be able to constrain $\sum m_{\nu}$ with higher precision and further explore a $\sum m_{\nu}$ scenario that can simultaneously reconcile high-precision cosmological observations with particle physics experimental results.

\section*{Acknowledgments}
This work was supported by the National Natural Science Foundation of China (Grants Nos.~12533001, 12575049, 12473001, 12305068, and 12305069), the National SKA Program of China (Grants Nos.~2022SKA0110200 and 2022SKA0110203), the China Manned Space Program (Grant No.~CMS-CSST-2025-A02), and the National 111 Project (Grant No.~B16009).

\bibliography{main} 

\end{document}